\documentstyle[epsf]{elsart}
\begin{document}
\newcommand{\ls}{\left(}
\newcommand{\rs}{\right)}
\newcommand{\beq}{\begin{equation}}
\newcommand{\eeq}{\end{equation}}
\newcommand{\beqa}{\begin{eqnarray}}
\newcommand{\eeqa}{\end{eqnarray}}
\newcommand{\bdm}{\begin{displaymath}}
\newcommand{\edm}{\end{displaymath}} 
\begin{frontmatter}
\title{ Constraints on the relativistic mean field of $\Delta$-isobar
in nuclear matter }
\author{D.S. Kosov$^{1}$, C. Fuchs$^{1}$, B.V. Martemyanov$^{1,2}$,
Amand Faessler$^{1}$
}
\address{
$^{1}$ Institut f\"ur Theoretische Physik, Universit\"at T\"ubingen,
Auf der Morgenstelle 14, D-72076 T\"ubingen, Germany
\\
$^{2}$ Institute for Theoretical and Experimental Physics,
B.Cheremushkinskaay 25,
117259 Moscow, Russia
}
\begin{abstract}
The effects of the presence of
$\Delta$-isobars in nuclear matter are studied in
the framework of relativistic
mean-field theory. The existence of stable nuclei at saturation density
imposes constraints on the  $\Delta$-isobar self-energy
and thereby
on the mean-field coupling constants of the  scalar
and vector mesons with $\Delta$-isobars. The range of possible values
for the scalar and vector coupling constants of $\Delta$-isobars
with respect to the nucleon coupling is investigated and
 compared
to recent predictions of QCD sum-rule calculations.
\end{abstract}
\end{frontmatter}

\newpage

The role of the $\Delta$-isobars in nuclear matter
has been discussed in various investigations in the
framework of effective hadron field theories, i.e. Quantum
Hadron Dynamics (QHD) \cite{Boguta82,Wehrberger89,Wehrberger93}.
In the framework of QHD \cite{Serot86} nucleons as well as
$\Delta$-isobars interact with
the surrounding nuclear medium by
the
exchange of effective scalar $\sigma$
and  vector $V^{\mu}$-mesons. However,
the coupling strengths of the respective
baryon-meson vertices are supposed to be different for nucleons and
deltas.
A lot of interesting physical observable
depend crucially on the choice of the respective coupling strengths
\cite{Boguta82,Wehrberger89,Wehrberger93}.
In the framework of effective QHD the
coupling constants for the nucleon-meson
vertices are effective quantities which are normally
adjusted to the nuclear matter saturation properties \cite{Serot86}.
They can, however, be also taken from different sources, e.g. from
chiral models \cite{furnstahl96} or from Dirac-Brueckner-Hartree-Fock
calculations \cite{fuchs95}.
There is up to now no definite information about
the scalar and vector delta coupling strengths.

Recent studies based on the
QCD finite density sum-rule \cite{Jin95} have shown that
the $\Delta$ vector self-energy is considerably weaker than that of the
nucleon, while its scalar self-energy is somewhat stronger.
However, the QCD sum-rule predictions for the scalar self-energy
are sensitive to the unknown density dependence of four-quark
condensates and due to this there is no reliable information about
the coupling constant of the $\Delta$-isobar with scalar mesons.
Thus phenomenological constraints on scalar and
vector coupling constants of $\Delta$-isobar are
of general interest, in particular because
they give new restrictions on the dependence of four-quark
condensates on the nucleon density. Further such considerations
show up the range for possible values of the respective vertices.
This is also of importance for phenomenological applications
such as, e.g., heavy ion collisions where the
decay and rescattering of $\Delta$-isobars are the prominent
sources of meson production \cite{metag93,RBUU}.
In relativistic heavy ion collisions the excess of deltas can reach
values
comparable to the saturation density of 
nuclear matter (so called
resonance matter) \cite{metag93}. In the theoretical description
within transport models the $\Delta$-isobars are commonly treated with
the same coupling strengths as the nucleons, i.e. the scalar and
vector components of the relativistic mean field are assumed
to be identical for nucleons and $\Delta$-isobars \cite{RBUU,fuchs97}.
This has of course influence on the delta dynamics as well as on
the description of inelastic collisions, e.g., the thresholds for
the production of mesons 
would be affected \cite{fuchs97}.
In order to use, e.g., pions or kaons as a source of
information on the hot and compressed
phase of nuclear matter the understanding of the properties
of $\Delta$-isobars in the nuclear medium appears to be indispensable.

In the present work we investigate the mixed phase of nuclear
and $\Delta$-matter within the framework of effective QHD.
The model system we are considering consists of nucleons $\Psi_{N}$,
$\Delta$-isobars which are treated as a Rarita-Schwinger
particle $\Psi_{\Delta \alpha}$ \cite{Rarita41}, effective
scalar $\sigma$ and vector $V^{\mu}$ meson fields  which are
both isoscalars. The inclusion of these mesons is
sufficient for the description of symmetric nuclear matter
with a vanishing isovector density and also in the mixed phase
where the total (zero) isospin should be conserved.
We adopt the extension of the
Walecka model \cite{Serot86} by including
$\Delta$-isobars as an additional degree of freedom in
the Lagrangian \cite{Boguta82}
\beqa
{\cal 
L} = {\cal L}_{f} -
 g_{v} \bar{\Psi}_{N} \gamma^{\mu} V_{\mu} \Psi_{N} &-&
g^{'}_{v} \bar{\Psi}^{\alpha}_{\Delta} \gamma^{\mu} V_{\mu}
\Psi_{\Delta \alpha} + \nonumber \\ 
g_{\sigma} \bar{\Psi}_{N}  \sigma \Psi_{N}
&+& g_{\sigma} \bar{\Psi}^{\alpha}_{\Delta}  \sigma \Psi_{\Delta \alpha }
+U(\sigma)
\quad .
\label{lagrang}
\eeqa
Here ${\cal L}_{f}$ collects a free Lagrangian of the
baryon fields $\Psi_{N}$, $\Psi_{\Delta \alpha}$ and the
$V^{\mu}$, $\sigma$ meson fields. The interaction terms
of
 the nucleons and $\Delta$-isobars with the meson fields are
explicitly given in Eq. (\ref{lagrang}). The respective coupling
constants   of nucleons $g_{\sigma}$ ($g_{v}$) and
 deltas $g^{'}_{\sigma}$ ($g^{'}_{v}$) with the scalar (vector)
mesons can be different in magnitude.
The phenomenological self-interaction of the scalar field $U(\sigma)$
specified as \cite{Boguta77}
\beq
U(\sigma)= -\frac{1}{3} B \sigma^{3} -\frac{1}{4} C \sigma^{4}
\label{nonlin}
\eeq
introduces an additional non-linear density dependence into the
model which improves the description of nuclear matter
bulk properties. The respective parameters taken
from Ref. \cite{nakai84} yield a saturation density
$\rho_0 = 0.167 fm^{-3}$, a binding energy of -15.8 MeV and
a compressibility of 290 MeV. To keep the model as
general as possible no restrictions
are imposed on the $\Delta$-meson vertices from the
beginning, i.e., $g^{'}_{\sigma}$ and $g^{'}_{v}$ are treated as
free
parameters. The range of their possible
 values is finally obtained
by the variation of $g^{'}_{\sigma , v}$ and the
consideration of the resulting equation of state.

In the mean field approximation
the meson fields in Eq. (\ref{lagrang}) are given by their classical
expectation values over the ground state of the nuclear system.
Thus one obtains the Lagrangian density in the mean-field approximation.
The Dirac field equations follow from mean field Lagrangian density as
\beqa
\left[ i \gamma^{\mu} \partial_{\mu}  -g_{v} \gamma^{\mu} V_{\mu}
-   M^{*}_{N} \right] \Psi_{N} (x) &=& 0
\label{dirac}
\\
\left[ i \gamma^{\mu} \partial_{\mu}
- g^{'}_{v} \gamma^{\mu} V_{\mu} -
   M^{*}_{\Delta} \right] \Psi^{\alpha}_{N} (x) &=& 0
\quad .
\label{dirac2}
\eeqa
The effective masses are given as
\beqa
M_{N}^{*} = M_{N} - g_{\sigma} \sigma \quad ,  \quad
M_{\Delta}^{*}= M_{\Delta} - g^{'}_{\sigma} \sigma
\label{mass}
\eeqa
where $M_{N}, M_{\Delta}$ stand for the bare nucleon
and $\Delta$-masses. In infinite nuclear matter
the mean field approximation
yields
the scalar $\sigma$ and vector
$V_{\mu} =\delta_{\mu 0} V_{0}$ meson fields which are
time and space independent
and directly proportional to the respective source terms, i.e.
\beqa
V_{0} &=& \frac{g_{v}}{m^{2}_{v}} \rho_{B}(N)
+ \frac{g^{'}_{v}}{m^{2}_{v}} \rho_{B}(\Delta)
\label{omega}
\\
m_{\sigma}^{2} \sigma + B \sigma^{2} +C \sigma^{3} &=&
g_{\sigma} \rho_{s}(N)
+ g^{'}_{\sigma} \rho_{s}(\Delta)
\label{sigma}
\eeqa
with $ \rho_{B}(N), \rho_{s}(N),\rho_{B}(\Delta), \rho_{s}(\Delta)$
being the nucleon and $\Delta$-isobar baryon and scalar densities.  The
equation for the effective nucleon mass $M_{N}^{*}$ is given by 
\beqa
M_{N}^{*}&=& M_{N} - \frac{g_{\sigma}^{2}}{m_{\sigma}^{2}} \rho_{s}(N)
-\frac{g_{\sigma} g^{'}_{\sigma} }{m_{\sigma}^{2}} \rho_{s}(\Delta)+ 
\nonumber  \\
&&\frac{B}{ g_{\sigma} m_{\sigma}^{2}} (M_{N}-M_{N}^{*})^{2}
+\frac{C}{ g^{2}_{\sigma} m_{\sigma}^{2}} (M_{N}-M_{N}^{*})^{3}
\label{mstar}
\eeqa
which has to be solved selfconsistently. The effective $\Delta$-mass
can be found as a function of $M_{N}^{*}$ and the scalar coupling
constant $g^{'}_{\sigma},g_{\sigma}$ from Eq. (\ref{mass}).


Chemical stability of the nucleons and $\Delta$-isobars requires that
the chemical potentials (Fermi energies) are equal:
\beq
E_{F} (N) = E_{F} (\Delta) \; ,
\label{equil}
\eeq
with
\beqa
E_{F}(N) &=& g_{v} V_{0}+\sqrt{k_{F}(N)^{2} +{M^{*}_{N}}^{2}}
\\
E_{F}(\Delta) &=& g^{'}_{v} V_{0}+\sqrt{k_{F}(\Delta)^{2} +
{M^{*}_{\Delta}}^{2}}
\quad .
\label{fermi}
\eeqa

The Fermi momentum $k_{F}(\Delta)$ of the $\Delta$-isobar matter 
is determined from Eq.(\ref{equil}) and $k_{F}(\Delta)$ is related to the 
density by usual relation, thereby
the $\Delta$-isobars density is governed by  
the requirement of a chemical equilibrium Eq.(\ref{equil}).

In Fig. 1  we show the equation of state, i.e., the energy per
baryon as a function of baryon density
for different ratios of scalar and vector
coupling constants of the $\Delta$ over those of
 the nucleon $r_s = g^{'}_{\sigma}/g_{\sigma}$
and $r_v = g^{'}_{v}/g_{v}$. It is found that
the presence of $\Delta$-isobars
becomes energetically favorable at higher densities ($\rho > 2\rho_{0}$)
where the mixed $\Delta$-nucleon phase appears. The
second minimum corresponds to new metastable state of nuclear
matter as already found in Ref. \cite{Boguta82}.

We obtain the first constraint on the delta coupling constants, i.e., on
ratios $r_s$ and $r_v$ by the requirement that the second
minimum should lie above the saturation energy
of normal nuclear matter (a). This means that in the mixed
$\Delta$-nucleon phase only a metastable state can occur.
An additional restriction on the coupling constant
can be obtained if one requires that
there are no $\Delta$-isobars present at saturation density. In other
words, the inclusion of a $\Delta$-isobar field in the
Lagrangian (\ref{lagrang}) should not effect the
value for the binding energy of nuclear matter
at saturation density, i.e. $k_{F}(\Delta) = 0$ at saturation
density (b). This requirement which is based
on the fact that there are no $\Delta$-isobars in the ground state of
finite nuclei imposes a "soft" restriction on the coupling constants
\cite{Wehrberger89}
\beq
r_s \leq 0.82 r_v + 
0.71
\quad .
\label{soft}
\eeq
Recent QCD finite density sum-rule calculations \cite{Jin95}
have shown that there exists
a larger net attraction for a $\Delta$-isobar than for a
nucleon in the nuclear medium.
Thus we can further restrict our consideration
to the cases where $r_{s} \ge 1$ and $r_v \le 1$ (c).

In Fig.2 we show the "window" for the possible values of
ratios which are allowed under the restrictions (a-c). The
constraints (a,b) (under the condition (c))
can thereby be summarized a
s
\beq
r_s \leq 1.01 r_v + 0.38
\label{hard}
\eeq
which imposes a more restrictive constraint on the ratios then
Eq. (\ref{soft}).

The QCD sum-rule calculations predict \cite{Jin95} that the
$\Delta$-isobar vector coupling is two times smaller (~$r_{v} \sim 0.4-0.5$)
than the corresponding nucleon coupling and
the magnitude of the  $\Delta$-isobar scalar coupling is
larger than that for nucleon (~$r_{s} \sim 1.3$). We checked these
 ratios between the $\Delta$-isobar and nucleon
coupling constants by our phenomenological
 model and we found in
this case that the depth of the second minimum is of more than one order
of
magnitude larger ($280 MeV$) than the minimum at saturation density.
This means that for the QCD predictions of Ref. \cite{Jin95}
the real ground state
of nuclear matter would occur to be a $\Delta$-isobar matter
at a density of about $\sim 3 \rho_{0}$ and
a binding energy of $\sim 280 MeV$.

To summarize we have investigated the properties of $\Delta$-isobars
in nuclear matter within the framework
 of Quantum Hadron Dynamics.
Based on the constraints that there are no $\Delta$-isobars
present at saturation density, the appearance of $\Delta$-isobars at
higher density leads to a metastable state which is not
the ground state of nuclear matter and that
the scalar interaction is more attractive
and the vector interaction is less repulsive for deltas
than for nucleons we have been able to derive a window
for possible values of mean field scalar and vector coupling constants
for the delta. It turned out that present QCD sum-rule
predictions do not fall into that window and seem to
yield too strong deviations of the $\Delta$-isobar self-energy
from the nucleon self-energy in the nuclear medium.

\begin{ack}
The authors acknowledge fruitful discussions
 with Drs. M.I. Krivoruchenko and
L. Sehn. B.V.M. is grateful to the Institute for Theoretical
Physics of University of Tuebingen for hospitality and financial
support. This work was partially supported 
by the Deutsche
Forschungsgemeinschaft (contract No FA67/20-1).
\end{ack}


\newpage
\begin{figure}[h]
\begin{center}
\leavevmode
\epsfxsize = 12cm
\epsffile[50 120 400 500]{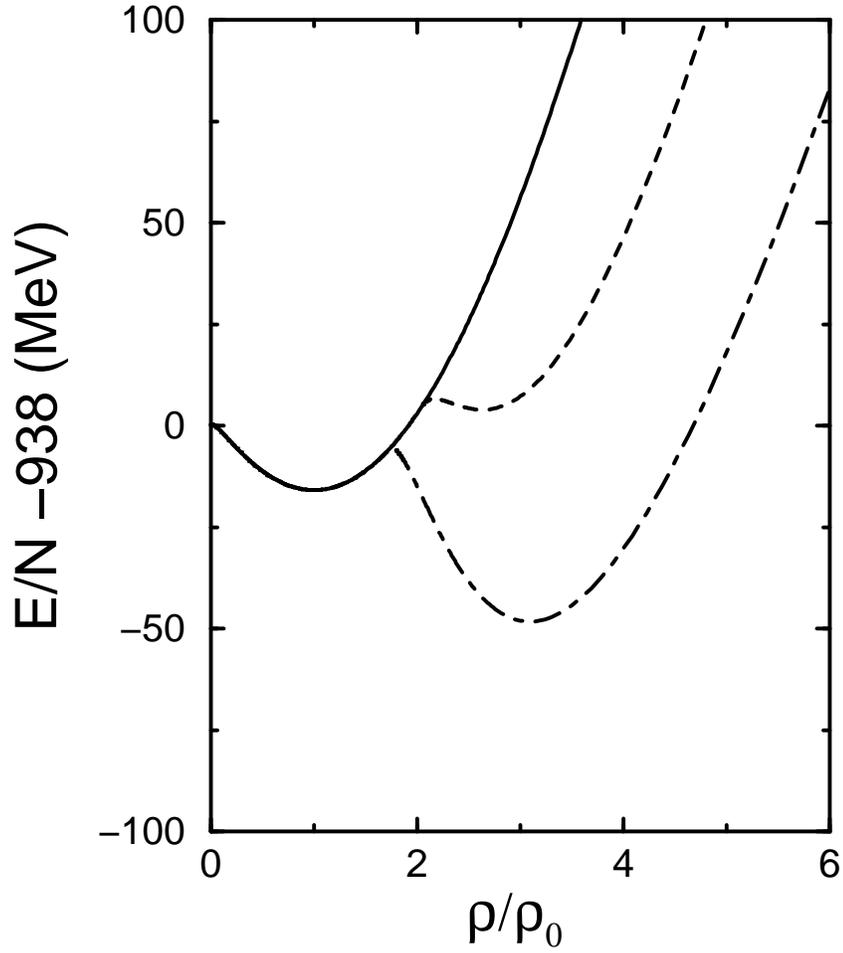}
\end{center}
\caption{
Equation of state, i.e.
binding energy per baryon versus density for nuclear matter
in the mixed nucleon and $\Delta$-isobar phase.
The calculation are performed
for various values of $r_s$ and $r_v$ ($r_s=1.35$, $r_v=1$  for the
dashed curve, and  $r_s=1.35$, $r_v=0.9$
for the dash-dotted curve). The solid curve shows
the nuclear matter equation of state without
$\Delta$-isobars.
}
\end{figure}
\begin{figure}[h]
\begin{center}
\leavevmode
\epsfxsize = 12cm
\epsffile[50 120 400 500]{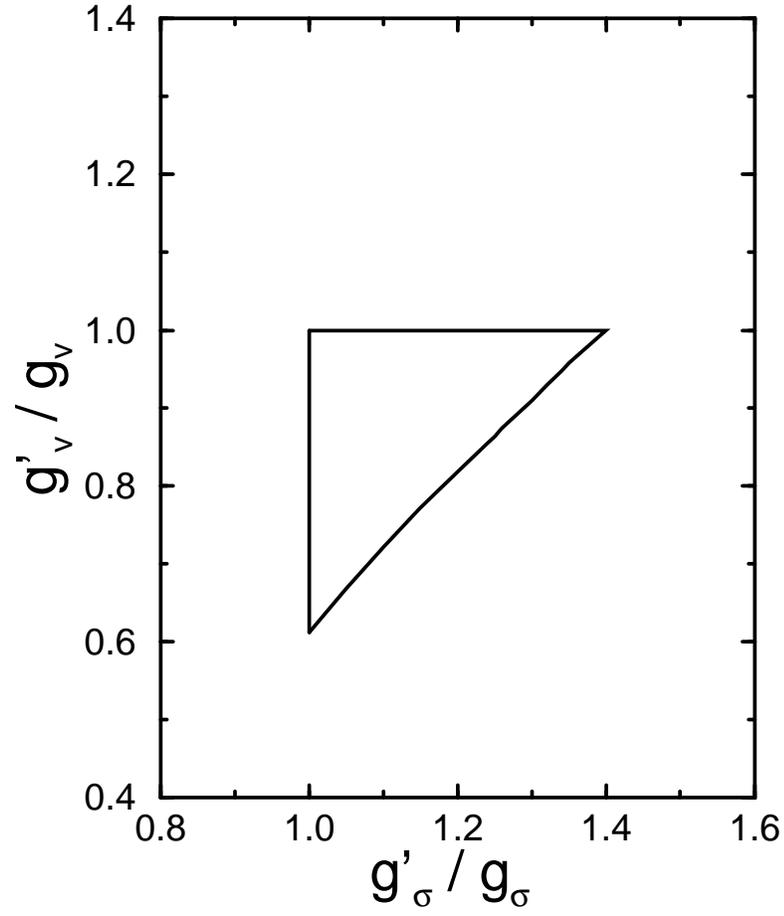}
\end{center}
\caption{
Values of ratios of scalar and vector coupling constant
for a $\Delta$-isobar to that for a nucleon
which are allowed under the assumption:
(a) The mixed $\Delta$-nucleon state is metastable, i.e. the
binding energy at the second minimum is smaller than in
ground state nuclear matter.
(b) There are no deltas present at the saturation density.
(c) The scalar mean field is more attractive
and the vector potential is less repulsive for deltas
than for nucleons.
}
\end{figure}
\end{document}